\documentclass[aps,twocolumn,showpacs]{revtex4}
\usepackage{graphicx}

\begin{document}

\title{The Fermi-Ulam Accelerator Model Under Scaling Analysis}

\author{Edson D.\ Leonel$^1$, P.V.E.\ McClintock$^1$ and
J. Kamphorst Leal da Silva$^2$}
\affiliation{
$^1$Department of Physics, Lancaster University, LA1 4YB, United Kingdom\\
$^2$Departamento de F\'{\i}sica, Instituto de Ci\^encias Exatas,
Universidade Federal de Minas Gerais, C. P. 702, CEP 30123-970, Belo
Horizonte, MG - Brazil
}

\date{\today} \widetext


\begin{abstract}
The chaotic low energy region of the Fermi-Ulam simplified
accelerator model is characterised by use of scaling analysis. It
is shown that the average velocity and the roughness (variance of
the average velocity) obey scaling functions with the same
characteristic exponents. The formalism is widely applicable,
including to billiards and to other chaotic systems.
\end{abstract}

\maketitle

Enrico Fermi \cite{ref1} attempted to describe cosmic ray
acceleration through a mechanism in which a charged particle can
be accelerated by collision with a time dependent magnetic field.
His original model was later modified and studied in different
versions, based on different approaches, one of which is the
well-known problem of a bouncing ball or Fermi-Ulam model (FUM)
\cite{ref3,ref4}. The main results for this version of the
problem, considering periodic oscillation, can be summarised as:
(i) it is described via the formalism of an area-preserving map;
(ii) it presents a set of invariant spanning curves in the phase
space for high energy; (iii) a set of KAM islands surrounded by a
chaotic sea can be observed in the low energy regime; and (iv)
small chaotic regions limited by two different invariant spanning
curves can be observed at intermediate energies. A related version
of this problem in a gravitational field, sometimes referred to as
a bouncer \cite{ref8}, presents a property, in contradistinction
to the bouncing ball problem, that depending on both initial
conditions and control parameters, the particle has unlimited
energy gain i.e.\ the basic condition needed for {\it Fermi
acceleration}. The difference between these apparently very
similar models was latter clarified by Lichtenberg et al
\cite{ref9}. The quantum problems corresponding to both the FUM
and bouncer models have also been investigated \cite{ref10,ref11,Referee}.
The special interest in studying these one-dimensional classical
systems arises because they allow direct comparison of theoretical
predictions with experimental results \cite{ref11a,ref11b}. 
Even more, the formalism used in the characterisation of such models
can immediately be extended to the so-called {\it billiards} class
of problems \cite{ref17,ref18,ref19}.

In this Letter, we characterise the average velocity, and its
variance which we will refer to as {\it roughness}, within the
chaotic sea of the phase space using scaling functions. One of our
tools, the roughness, is an extension of the formalism used to
characterise rough surfaces \cite{barabasi} which, as we will
show, is immediately applicable to chaotic orbits in the problem
of time dependent potential wells \cite{ref12,ref13,ref14,ref15}
as well as to billiards problems \cite{ref16,ref16a}. This scaling
scenario represents the first characterisation of the
integrability-chaos transition in this problem and should be 
applicable to several billiard problems. The
formalism may therefore prove useful in characterising classes of
universality.

Let us describe the system and how to characterise its dynamical
evolution. It consists of a classical particle bouncing between
two rigid walls, one of which is fixed; the other moves
periodically in time with a normalised amplitude $\epsilon$. We
will describe the system using a map
$T(V_n,\phi_n)=(V_{n+1},\phi_{n+1})$ which gives the new velocity
of the particle and the corresponding phase of the moving wall
immediately after the particle suffers a collision with it. We
will use a simplification \cite{ref20} in our description of this
problem: we will suppose that both walls are fixed but that, when
the particle suffers a collision with one of the walls, it
exchanges momentum as if the wall were moving. This simplification
carries the huge advantage of allowing us substantially to speed
up our numerical simulations compared with the full model. It is
usefully applicable because the main dynamical properties of the
system are preserved under such conditions. Incorporating this
simplification in the model and using dimensionless variables, the
map is written as \cite{ref3}
\begin{equation}
T:\left\{\begin{array}{ll}
V_{n+1}=|V_n-2\epsilon\sin(\phi_{n+1})|~~\\
\phi_{n+1}=\phi_n+{2\over{V_n}}~~~ {\rm mod} 2\pi\\
\end{array}
\right..
\label{eq1}
\end{equation}
The term $2/V_n$ specifies the length of time during which the
particle travels between collisions, while
$-2\epsilon\sin(\phi_{n+1})$ gives the corresponding fraction of
velocity gained or lost in the collision. The modulus function is
introduced to avoid the particle leaving the region between the
walls. We stress that the approximation of using the simplified
FUM is valid in the limit of small $\epsilon$. So, the transition
from integrability ($\epsilon=0$) to chaos ($\epsilon\ne 0$),
characterising the birth of the chaotic sea, can be well
described.

\begin{figure}[b]
\centerline{\includegraphics[width=0.93\linewidth]{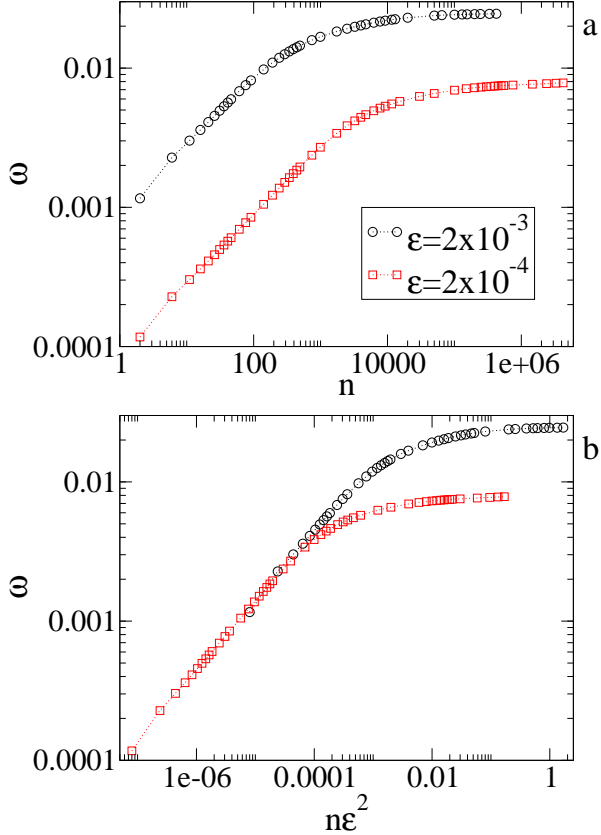}}
\caption{(a) Behaviour of the roughness $\omega$ as a function of
the iteration number $n$. (b) Behaviour of $\omega$ as a function
of $n\epsilon^2$. Both curves were derived from an ensemble
average of $50,000$ different initial conditions starting with
$V_0\approx 0$.} \label{fig1}
\end{figure}

We will concentrate on the scaling behaviour present in
the chaotic sea. We investigate the evolution of the velocity
averaged in $M$ initial phases, namely
\begin{equation}
V(n,\epsilon,V_0)={1\over M}\sum_{j=1}^M V_{n,j}~~,
\label{eq_jaff_1}
\end{equation}
where $V_0$ is the initial velocity and $j$ refers to a
sample of the ensemble. In order to define the roughness
\cite{barabasi}, we first consider the average of velocity over
the orbit generated from one initial phase
\begin{equation}
\overline{V}(n,\epsilon,V_0)={1\over{n}}\sum_{i=0}^n V_i~.
\end{equation}
We then evaluate the interface width around this averaged
velocity. Finally, the roughness is defined by considering an
ensemble of $M$ different initial phases:
$$
\omega(n,\epsilon,V_0)\equiv{1\over{M}}\sum_{j=1}^M\left[
\sqrt{\overline{V^2}_j(n,\epsilon,V_0)-\overline{V}_j^2(n,\epsilon,V_0)}~\right].
$$
Fig.\ \ref{fig1} shows the behaviour of the
roughness for two different control parameters.
We can see in Fig.\ \ref{fig1}(a) that the roughness grows for
small iteration number $n$ and then saturates at large $n$. The
change from growth to saturation is characterised by a crossover
iteration number $n_x$.
\begin{figure}[t]
\centerline{\includegraphics[width=0.95\linewidth]{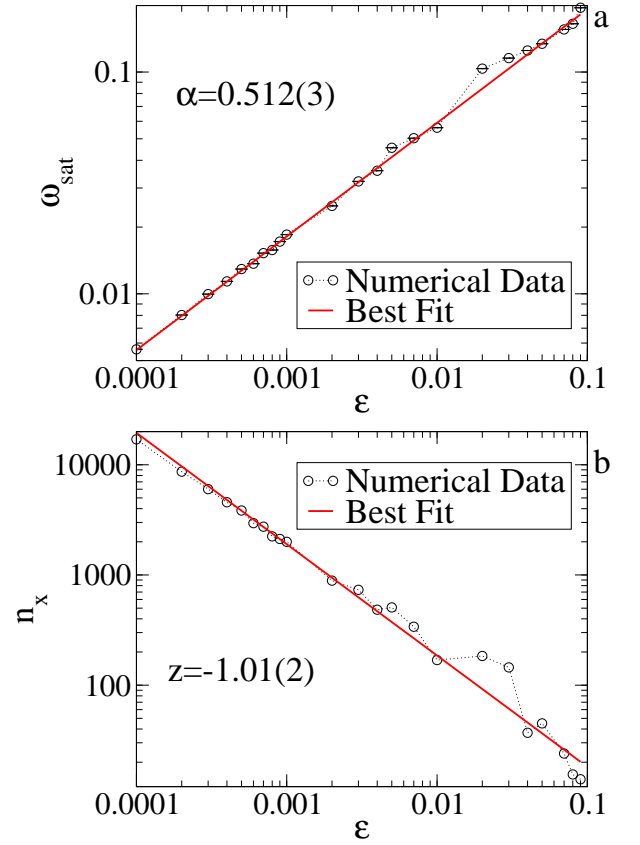}}
\caption{(a) Plot of $\omega_{\rm sat}$ against the control
parameter $\epsilon$. (b) The crossover iteration number $n_x$ as
a function of $\epsilon$. } \label{fig2}
\end{figure}
\begin{figure}[b]
\centerline{\includegraphics[width=1.0\linewidth]{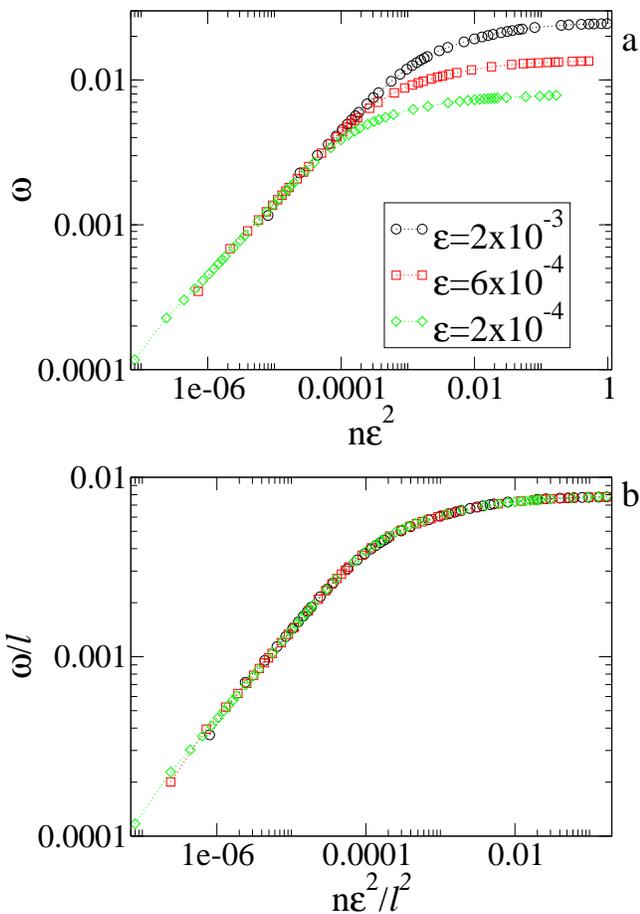}}
\caption{(a) Roughness evolution for different values of the
control parameter $\epsilon$. (b) Collapse of the curves from (a)
onto a universal curve. Both (a) and (b) were obtained using
$V_0\approx 0$.} \label{fig3}
\end{figure}
\begin{figure}[t]
\centerline{\includegraphics[width=1.0\linewidth]{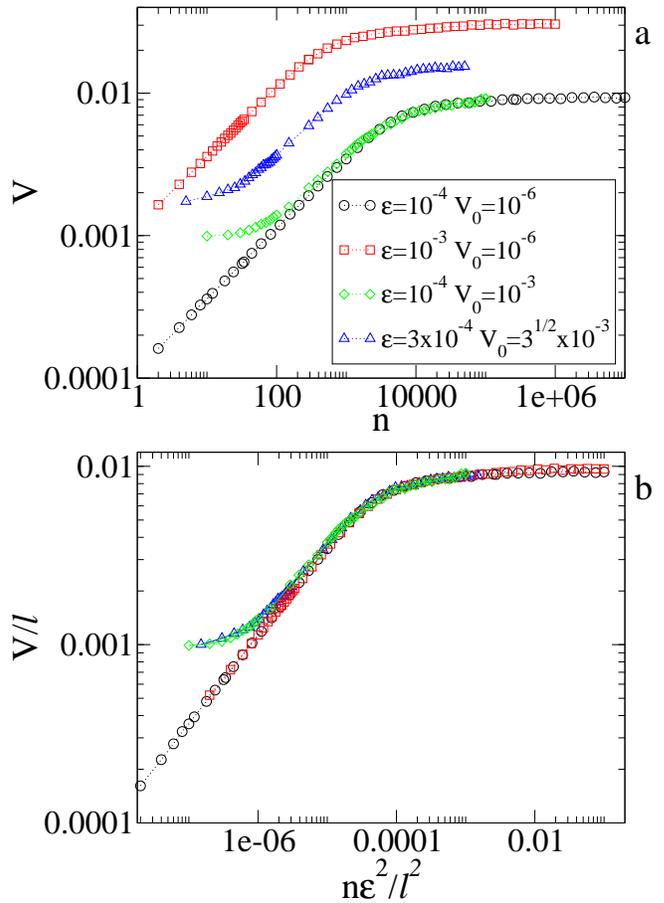}}
\caption{Behaviour of the average velocity $V$ as a function of
$n$ for different values of $\epsilon$ and $V_0$. (a) The original
time series. (b) Collapse of the data onto a universal curve.}
\label{fig4}
\end{figure}
We can also see that different values of the control parameter
$\epsilon$ generate different behaviours for short $n$. This
indicates that $n$ is not a scale variable. However, it turns out
that the transformation $n\rightarrow n\epsilon^2$ coalesces the
curves for small $n$ as illustrated in Fig.\ \ref{fig1}(b). We
therefore infer: (i) that for $n \ll n_x$ the roughness grows
according to
\begin{equation}
\omega(n\epsilon^2,\epsilon,V_0)\propto ({n\epsilon^2})^{\beta}~,
\label{eq5}
\end{equation}
where exponent $\beta$ is called the growth exponent; (ii) that,
as the iteration number increases, for $n \gg n_x$, the roughness
reaches a saturation regime that is describable as
\begin{equation}
\omega_{\rm sat}(\epsilon)\propto \epsilon^{\alpha}~, \label{eq6}
\end{equation}
where $\alpha$ is the roughening exponent; and (iii) that the
crossover iteration number $n_x$ marking the approach to
saturation is
\begin{equation}
n_x(\epsilon,V_0)\propto\epsilon^z~,
\label{eq7}
\end{equation}
where $z$ is called the dynamical exponent. With these initial
suppositions, we can now describe the roughness formally in terms
of a scaling function,
\begin{equation}
\omega(n\epsilon^2,\epsilon,V_0)=l\omega(l^an\epsilon^2,l^b\epsilon,l^cV_0)~,
\label{eq8}
\end{equation}
where $l$ is the scaling factor and  $a$, $b$, and $c$ are
referred to as scaling dimensions. It is important to stress that
these scaling dimensions $a$, $b$ and $c$ must be related to the
characteristic exponents $\alpha$, $\beta$ and $z$. All of the
above discussion is also valid for the average velocity $V$. To
relate the exponent $\beta$ to the scaling dimensions, we  chose
$l=(n\epsilon^2)^{-{1\over{a}}}$. This allows us to rewrite
(\ref{eq8}) as
\begin{equation}
\omega(n\epsilon^2,\epsilon,V_0)=(n\epsilon^2)^{-{1\over{a}}}
\omega_1\left((n\epsilon^2)^{-{b\over{a}}}\epsilon,
(n\epsilon^2)^{-{c\over{a}}}V_0\right)~.
\label{eq9}
\end{equation}
The function
$\omega_1=\omega\left(1,(n\epsilon^2)^{-{b\over{a}}}\epsilon,
(n\epsilon^2)^{-{c\over{a}}}V_0\right)$ is supposed constant for
$n \ll n_x$. Comparing equations (\ref{eq9}) and (\ref{eq5}), we
can however conclude that $-{1\over{a}}=\beta$. Choosing
$l=\epsilon^{-{1\over{b}}}$, we have that
\begin{equation}
\omega(n\epsilon^2,\epsilon,V_0)=\epsilon^{-{1\over{b}}}
\omega_2\left((n\epsilon^2)\epsilon^{-{a\over{b}}},
\epsilon^{-{c\over{b}}}V_0\right)~,
\label{eq10}
\end{equation}
where $\omega_2=\omega\left(
(n\epsilon^2)\epsilon^{-{a\over{b}}},1,
\epsilon^{-{c\over{b}}}V_0\right)$ is assumed constant for $n \gg
n_x$. Comparison of equations (\ref{eq10}) and (\ref{eq6}) shows
that $-{1\over{b}}=\alpha$. It is less straightforward to obtain
the exponent $c$. To do so, we use a result from a recent paper
where two of us \cite{ref23} utilised a connection with the well
known Standard Model (SM) \cite{ref3} to describe the position of
the first invariant spanning curve (FISC) above the chaotic sea in
the FUM. It was shown that the control parameter $\epsilon$ could
be related to a typical mean velocity on the FISC in the FUM to
give an effective control parameter $K_{\rm
eff}=4\epsilon/{V^*}^2\approx 0.97\ldots$ which is gratifyingly
close to the value of the control parameter 
$K_{\rm c}=0.9716\ldots$ at which the SM exhibits a transition
from local to globally stochastic behaviour \cite{ref24}. We can
thus rewrite the effective control parameter $K_{\rm eff}$ in
terms of scaled variables as
\begin{equation}
K_{\rm eff}={4(l^b\epsilon)\over{(l^cV_0)^2}}=
{4\epsilon\over{V_0^2}}{l^b\over{l^{2c}}}~. \label{eq11}
\end{equation}
which implies that $b-2c=0$. Using our result for the exponent
$b$, we find $c=-{1\over{2\alpha}}$. Note that all scaling
dimensions are therefore determined if we can obtain the exponents
$\alpha$ and $\beta$ numerically. The exponent $\alpha$ is
obtained in the asymptotic limit of large iteration number and it
is independent of $V_0$. Fig.\ \ref{fig2}(a) illustrates an
attempt to characterise this exponent using the extrapolated
saturation roughness. Extrapolation is required because, even
after $10^3 n_x$ iterations, the roughness has still not quite
reached saturation. From a power law fit, we obtain
$\alpha=0.512(3)\approx 1/2$. We can thus rewrite Eq.\ (\ref{eq8})
as
\begin{equation}
\omega(n\epsilon^2,\epsilon,V_0)=
(n\epsilon^2)^{\beta}g\left[(n\epsilon^2)^{-2\beta}\epsilon,
(n\epsilon^2)^{-\beta}V_0\right]~.
\label{eq12}
\end{equation}
In order to obtain $\beta$, we observe that we have two ``time''
scales in equation (\ref{eq12}), namely $n_x^{\prime}$ and
$n_x^{\prime\prime}$ and that the second one
($n_x^{\prime\prime}$) is basically zero if we chose $V_0\approx
0$. Then we determine $\beta$ from the short ``time'' behaviour
($n \ll n_x^{\prime}$). After averaging over different values of
the control parameter $\epsilon$ in the range $\epsilon\in
[10^{-4},10^{-1}]$, we then obtain $\beta=0.496(6)\approx 1/2$.
Therefore, the scaling dimensions describing the scaling of the
chaotic sea in the limit of small $\epsilon$ are $a=b=-2$ and
$c=-1$. From Eqs.\ (\ref{eq7}) and (\ref{eq9}) we find that the
scaling relation for the exponent $z$ is $z=\alpha/\beta-2$.
Considering the previous values of both $\alpha$ and $\beta$, we
obtain that $z=-1$. The exponent $z$ can be also obtained
numerically. Fig.\ \ref{fig2}(b) shows the behaviour of the
crossover iteration number $n_x$ as function of the control
parameter $\epsilon$. The power law fit gives us that
$z=-1.01(2)$, in good accord with the scaling result. The scaling
for $V_0\approx 0$ is demonstrated in Fig.\ \ref{fig3}, where the
three different curves for the roughness in (a) are very well
collapsed onto the universal curve seen in (b) when we normalise
the quantities with  $a=b=-2$.

The case with initial velocity $V_0\ne 0$ is better illustrated by
the average velocity (see Fig.\ \ref{fig4}). Now, we must consider
two ``time'' scales, namely $n_x^{\prime}\propto 1/\epsilon$ and
$n_x^{\prime\prime}\propto V_0^2/\epsilon^2$. From  equation
(\ref{eq11}) (see also Ref.\ \cite{ref23}), the maximum initial
velocity inside the chaotic sea is $V_{0,{\rm max}}\approx
2\epsilon^{1/2}$ implying that the second time scale has a maximum
value of ($n_x^{\prime\prime}\sim 4n_x^{\prime}$). So we observe
that two different kinds of behaviour may occur, for
$n_x^{\prime\prime}<n_x^{\prime}$ or $n_x^{\prime\prime}\sim
n_x^{\prime}$. When $V_0=10^{-6}$, we have
$n_x^{\prime\prime}\approx 0$ and we can see in Fig.\
\ref{fig4}(a) that the curves for $\epsilon=10^{-4}$ and
$\epsilon=10^{-3}$ show only two regimes: (1) a growth in power
law for $n \ll n_x^{\prime}$ and (2) the saturation regime for $n
\gg n_x^{\prime}$. Considering $V_0=10^{-3}$ and
$\epsilon=10^{-4}$ we have that $n_x^{\prime\prime}<n_x^{\prime}$
and we can see for such curve in Fig.\ \ref{fig4}(a) three
regimes. For $n\ll n_x^{\prime\prime}$, the average velocity is
basically constant. When $n_x^{\prime\prime}<n<n_x^{\prime}$, the
curve growth and begin to follow the curve of $V_0=10^{-6}$ and
same $\epsilon$. In this window of $n$, we have a growth with a
smaller effective exponent $\beta$. Finally, for $n\gg
n_x^{\prime}$ we have the saturation regime. It is shown in Fig\
\ref{fig4}(b) that the collapse of the curves holds even for
$V_0\not=0$, implying that the inferred scaling form
$V(n\epsilon^2,\epsilon,V_0)$ with exponents $a=b=-2$ and $c=-1$
is also correct.

In summary, we have characterised the average velocity and its
variance (roughness) in the chaotic sea in the simplified FUM by
use of a scaling function. We show that the critical exponents
$\beta$, $\alpha$ and $z$ are connected by a scaling relation. We
emphasise that this behaviour is valid for small values of
$\epsilon$ and it is immediately extendable to other average
quantities. We have characterised, for the first time,  the
scaling appearing in the integrability$\rightarrow$chaos
transition (from $\epsilon=0$ to $\epsilon\not=0$) of the FUM. The
scaling scenario should also hold for billiard systems, so that
this kind of formalism should be useful for characterising
asymptotic properties in such problems. It should be possible to
extend it to encompass time-dependent Hamiltonian systems and a
huge class of other problems exhibiting chaotic behaviour.

E. D. Leonel was supported by Conselho Nacional de
Desenvolvimento Cient\'{\i}fico CNPq, from Brazil. The numerical
results were obtained in the Centre for High Performance Computing
in Lancaster University. The work was supported in part by the
Engineering and Physical Sciences Research Council (UK).
J. K. L. da Silva thanks to Conselho Nacional de Desenvolvimento
Cient\'{\i}fico (CNPq) and Funda\c c\~ao de Amparo a Pesquisa de
Minas Gerais (Fapemig), Brazilian agencies.


\begin{thebibliography}{99}

\bibitem{ref1} E. Fermi, Phys. Rev. {\bf 75}, 1169 (1949).

\bibitem{ref3} A. J. Lichtenberg, M.A. Lieberman. {\it Regular
and Chaotic Dynamics}(Appl. Math. Sci. {\bf 38}, Springer Verlag,
New York, 1992).

\bibitem{ref4} M.A. Lieberman and A. J. Lichtenberg. Phys. Rev. A
{\bf 5}, 1852 (1971).

\bibitem{ref8} L. D. Pustilnikov, Theor. Math. Phys. {\bf 57},
1035 (1983); L. D. Pustilnikov, Sov. Math. Dokl. {\bf 35(1)}, 88
(1987); L. D. Pustilnikov, Russ. Acad. Sci. Sb. Math. {\bf 82(1)},
 231 (1995).

\bibitem{ref9} A. J. Lichtenberg, M.A. Lieberman and R. H. Cohen
Physica D {\bf 1}, 291 (1980).

\bibitem{ref10} G. Karner, J. Stat. Phys. {\bf 77}, 867 (1994).

\bibitem{ref11} S. T. Dembinski, A. J. Makowski and P. Peplowski,
Phys. Rev. Lett. {\bf 70}, 1093 (1993).

\bibitem{Referee} J. V. Jos\'e and R. Cordery, Phys. Rev. Lett. 
{\bf 56}, 290 (1986).

\bibitem{ref11a} Z. J. Kowalik, M. Franaszek and P.
Pieranski, Phys. Rev. A {\bf 37}, 4016 (1988).

\bibitem{ref11b}  S. Warr, W. Cooke, R. C. Ball and J. M. Huntley,
Physica A {\bf 231}, 551 (1996).

\bibitem{ref17} N. Sait\^o, H. Hirooka, J. Ford, F. Vivaldi and G.
 H. Walker, Physica D {\bf 5}, 273 (1982).

\bibitem{ref18} E. Canale, R. Markarian, S. O. Kamphorst and
S. P. de Carvalho, Physica D {\bf 115}, 189 (1998).

\bibitem{ref19} A. Loskutov, A. B. Ryabov and L. G. Akinshin,
J. Phys. A: Math. Gen. {\bf 33}, 7973 (2000).

\bibitem{barabasi} A. -L. Barab\'asi, H. E. Stanley, {\it Fractal Concepts in
Surface Growth} (Cambridge University Press, Cambridge, 1985).

\bibitem{ref12} J. L. Mateos and J. V. Jos\'e, Physica A
{\bf 257}, 434 (1998).

\bibitem{ref13} J. L. Mateos, Phys. Lett. A {\bf 256}, 113 (1999).

\bibitem{ref14} G. A. Luna-Acosta, G. Orellana-Rivadeneyra, A.
Mendoza-Galv\'an and C. Jung, Chaos, Solitons and Fractals {\bf
12}, 349 (2001).

\bibitem{ref15} E. D. Leonel and J. K. L. da Silva, Physica A
{\bf 323}, 181 (2003).

\bibitem{ref16} M. V. Berry, Eur. J. Phys. {\bf 2}, 91 (1981).

\bibitem{ref16a} M. Robnik and M. V. Berry, J. Phys. A {\bf 18}, 1361 (1985). 

\bibitem{ref20} The simplified FUM was introduced by
Lichtenberg and Lieberman and can be found e.g.\ in Ref. \cite{ref3}.

\bibitem{ref23} E. D. Leonel, J. K. L. da Silva and S. O.
Kamphorst, Physica A {\bf 331}, 435 (2004).

\bibitem{ref24} We have used the same notation as the original work
\cite{ref3} although, at that time, stochasticity was frequently
referred as to chaotic behaviour. However we stress that the
transition is from locally to globally chaotic behaviour. In the
FUM, we mean by {\it locally} that the chaotic
behaviour is confined by two different invariant spanning curves.

\end{thebibliography}
\end{document}